\documentclass[sn-mathphys]{sn-jnl}%
\jyear{2021}%
\theoremstyle{thmstyleone}%
\theoremstyle{thmstyletwo}%
\theoremstyle{thmstylethree}%
\begin{document}

\title[ ]{Coherent X-ray Spectroscopy Elucidates Nanoscale Dynamics of Plasma-Enhanced Thin-Film Growth}

\author*[1,5]{\fnm{Peco} \sur{Myint}}\email{pmyint@anl.gov}
\equalcont{These authors contributed equally to this work.}
\author*[2]{\fnm{Jeffrey M.} \sur{Woodward}}\email{jeffrey.woodward@nrl.navy.mil}
\equalcont{These authors contributed equally to this work.}
\author*[1]{\fnm{Chenyu} \sur{Wang}}\email{cywang96@bu.edu}
\equalcont{These authors contributed equally to this work.}
\author[3]{\fnm{Xiaozhi}
\sur{Zhang}}
\author[4]{\fnm{Lutz}
\sur{Wiegart}}
\author[4]{\fnm{Andrei}
\sur{Fluerasu}}
\author[3]{\fnm{Randall L.}
\sur{Headrick}}
\author[2,6]{\fnm{Charles R.}
\sur{Eddy, Jr.}}
\author*[1]{\fnm{Karl F.}
\sur{Ludwig}}\email{ludwig@bu.edu}

\affil[1]{\orgdiv{Department of Physics and Division of Materials Science and Engineering}, \orgname{Boston University}, \orgaddress{ \city{Boston}, \state{MA}, \country{USA}}}

\affil[2]{\orgdiv{Electronics Science and Technology Division}, \orgname{U.S. Naval Research Laboratory}, \orgaddress{\city{Washington DC}, \country{USA}}}

\affil[3]{\orgdiv{Department of Physics and Materials Science Program}, \orgname{ University of Vermont}, \orgaddress{\city{Burlington}, \country{USA}}}

\affil[4]{\orgdiv{National Synchrotron Light Source II}, \orgname{Brookhaven National Laboratory}, \orgaddress{\city{Upton}, \country{USA}}}

\affil[5]{\orgdiv{Current address: X-ray Science Division}, \orgname{Argonne National Laboratory}, \orgaddress{ \city{Lemont}, \state{IL}, \country{USA}}}

\affil[6]{\orgdiv{Current address: ONR Global}, \orgaddress{ \city{Ruislip}, \country{UK}}}

\abstract{Sophisticated thin film growth techniques increasingly rely on the addition of a plasma component to open or widen a processing window, particularly at low temperatures.  However, the addition of the plasma into the growth environment also complicates the surface dynamical evolution.  Taking advantage of continued increases in accelerator-based X-ray source brilliance, this real-time study uses X-ray Photon Correlation Spectroscopy (XPCS) to elucidate the nanoscale surface dynamics during Plasma-Enhanced Atomic Layer Deposition (PE-ALD) of an epitaxial indium nitride film.  XPCS examines the evolution of the coherent X-ray scattering speckle pattern, which is a fingerprint of the unique sample microstructure at each moment in time.   In PE-ALD, ultrathin films are synthesized from repeated cycles of alternating self-limited surface reactions induced by temporally-separated pulses of material precursor and plasma reactant, allowing the influence of each on the evolving morphology to be examined.  During the heteroepitaxial 3D growth examined here, sudden changes in surface structure during initial film growth, consistent with numerous overlapping stress-relief events, are observed. When the film becomes continuous, the nanoscale surface morphology abruptly becomes long-lived with correlation time spanning the period of the experiment. Throughout the growth experiment, there is a consistent repeating pattern of correlations associated with the cyclic growth process, which is modeled as transitions between different surface states.  The plasma exposure does not simply freeze in a structure that is then built upon in subsequent cycles, but rather there is considerable surface evolution during all phases of the growth cycle.}


\maketitle
\newpage
\section*{Introduction}\label{sec1}

Societal needs for improved materials in energy \cite{gupta2022recent, he2019thin}, communications  \cite{cheng2022emerging, liu2020two} and computing \cite{de2021materials, liu20192d} continue to drive the development of high-performance thin films.  This has led to a progression of increasingly sophisticated film growth approaches, including increasing incorporation of a plasma component.  Application of a plasma during growth can open or widen a process window, particularly enabling low-temperature growth in which energy from energetic species compensates for lack of thermal energy. \cite{profijt2011plasma, knoops2019status, boris2020role}  However, the details of plasma-based growth processes are seldom fully understood. Understanding the processes driving growth in a plasma environment, as well as in other complex growth methods, needs correspondingly innovative methods to reveal the detailed nano- and atomic-scale surface structural evolution.

For studies of sophisticated growth approaches, real-time investigations have natural advantages.  They provide a complete temporal record of kinetics processes and there is no concern about sample changes in the period between the end of processing and a {\it post facto} measurement.  This attribute is especially important for the study of high temperature surfaces on which relaxation can be relatively rapid. Moreover, real-time grazing-incidence X-ray experiments allow penetration of complex ambient environments and control over surface/bulk sensitivity by variation of the incident and/or exit angle \cite{perret2014real, fujii1996situ, perret2017island, gupta2021situ, yan2020situ, fuoss1989atomic, fuoss1992time, stephenson1999observation}.  However, traditional “low-coherence” X-ray scattering is only sensitive to average structural information.  Thus, for instance, such an experiment can determine that the film morphology is statistically unchanging during a particular part of the growth process.  But it cannot itself determine the degree of nanoscale conformality of the growth. In contrast, the continuing increase in accelerator-based X-ray source brilliance has enabled the development of approaches measuring the evolution of the coherent X-ray scattering speckle pattern, which is a fingerprint of the unique sample microstructure at each moment in time \cite{leheny2012xpcs}.  However, application of coherent X-ray scattering in a surface-sensitive mode during growth has significant experimental and analysis challenges and the approach is still in the early stages of development
\cite{rainville2015,ulbrandt2016coherent,headrick2019coherent,ju2019coherent,wang2022amorphous}.  

In the present work, we exhibit the ability to use the coherent x-ray scattering technique of X-ray Photon Correlation Spectroscopy (XPCS) to elucidate sub-monolayer structural dynamics during a pulsed plasma-driven self-limited growth process.  Self-limiting pulsed growth processes, including the prototypical example of Atomic Layer Deposition (ALD), offer unique advantages \cite{oviroh2019new, george2010atomic}. These attributes have made ALD a growth method of choice for the deposition of amorphous high-$\kappa$ dielectric layers in the semiconductor industry \cite{johnson2014brief, leskela2003atomic, gougousi2016atomic, liu2018overview, mollah2019ultra, karabulut2018electrical} and interest continues to build in applying it to the growth of epitaxial materials, particularly nitrides \cite{nepal2013epitaxial, nepal2017real, feng2018epitaxial, nepal2019understanding, woodward2019influence, obrien2020, hsu2020direct, woodward2022influence} and oxides \cite{klepper2007epitaxial, kukushkin2016epitaxial, schuisky2002epitaxial,wheeler2020phase}. In Plasma-Enhanced ALD (PE-ALD), the processes of material deposition and crystallization are temporally separated in a cyclic procedure, allowing the influence of each on evolving morphology to be examined.  However, it’s known that the simple ALD model of uniform growth across a surface with each growth cycle is inaccurate. In fact, mounding is often observed, as in other growth processes \cite{puurunen2004island}.  It’s clear that our fundamental understanding of the nanoscale dynamical processes occurring during such plasma-driven self-limiting growth is inadequate, inhibiting our ability to develop new growth methods for complex materials systems. Here we show that surface-sensitive XPCS provides a new window into the dynamics of such plasma-driven pulsed growth processes.

\section*{Experimental setup}\label{experimental_setup}

\begin{figure}[h!]
\centering\includegraphics[width=9cm]{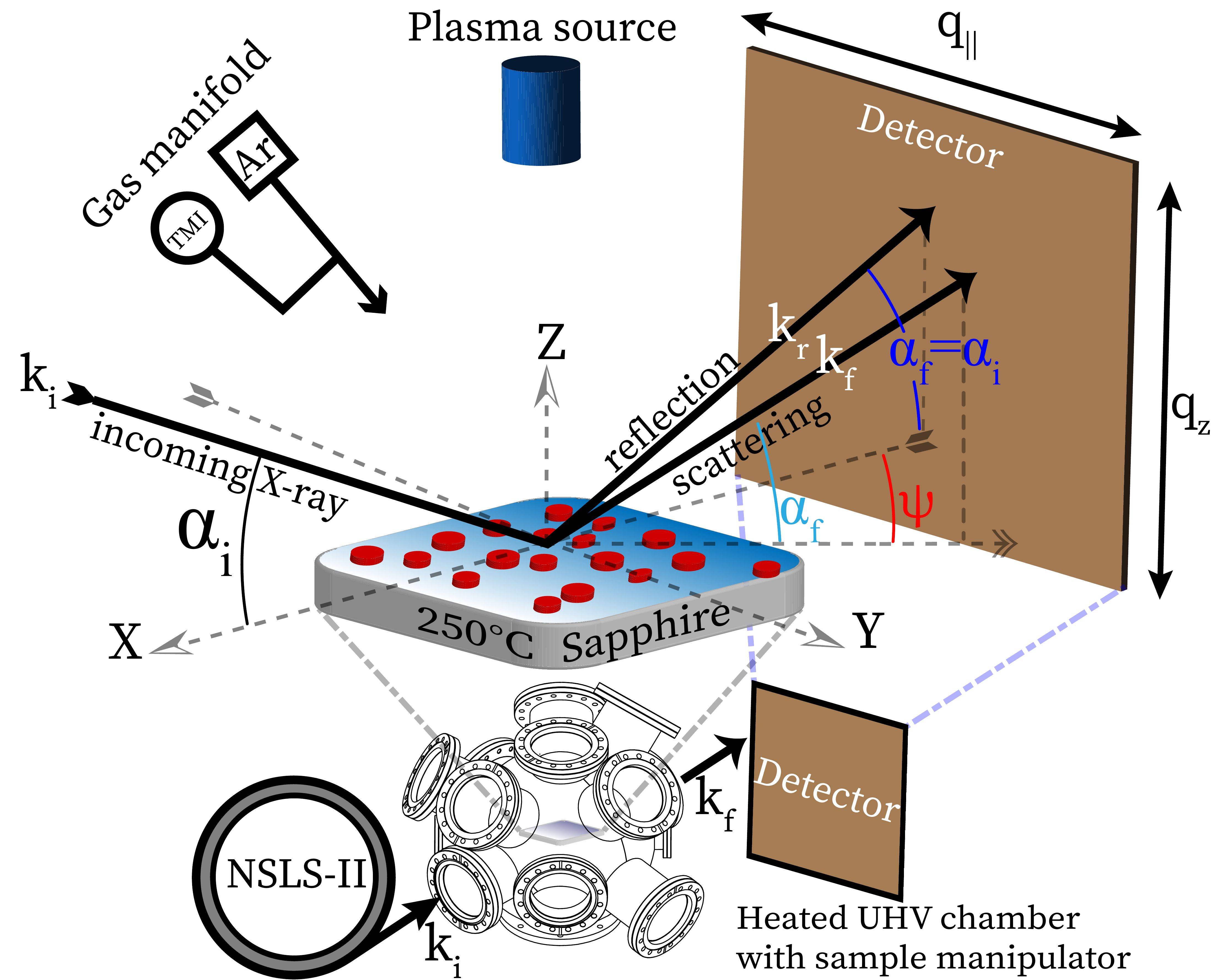}
\caption{Schematic diagram of PE-ALD growth real-time XPCS study. X-rays enter the custom-made chamber where the PE-ALD process is contained. Scattered X-rays from the changing surface are recorded by an area detector.}
\label{fig:setup}
\end{figure}

\begin{figure}[h!]
\centering\includegraphics[width=5cm]{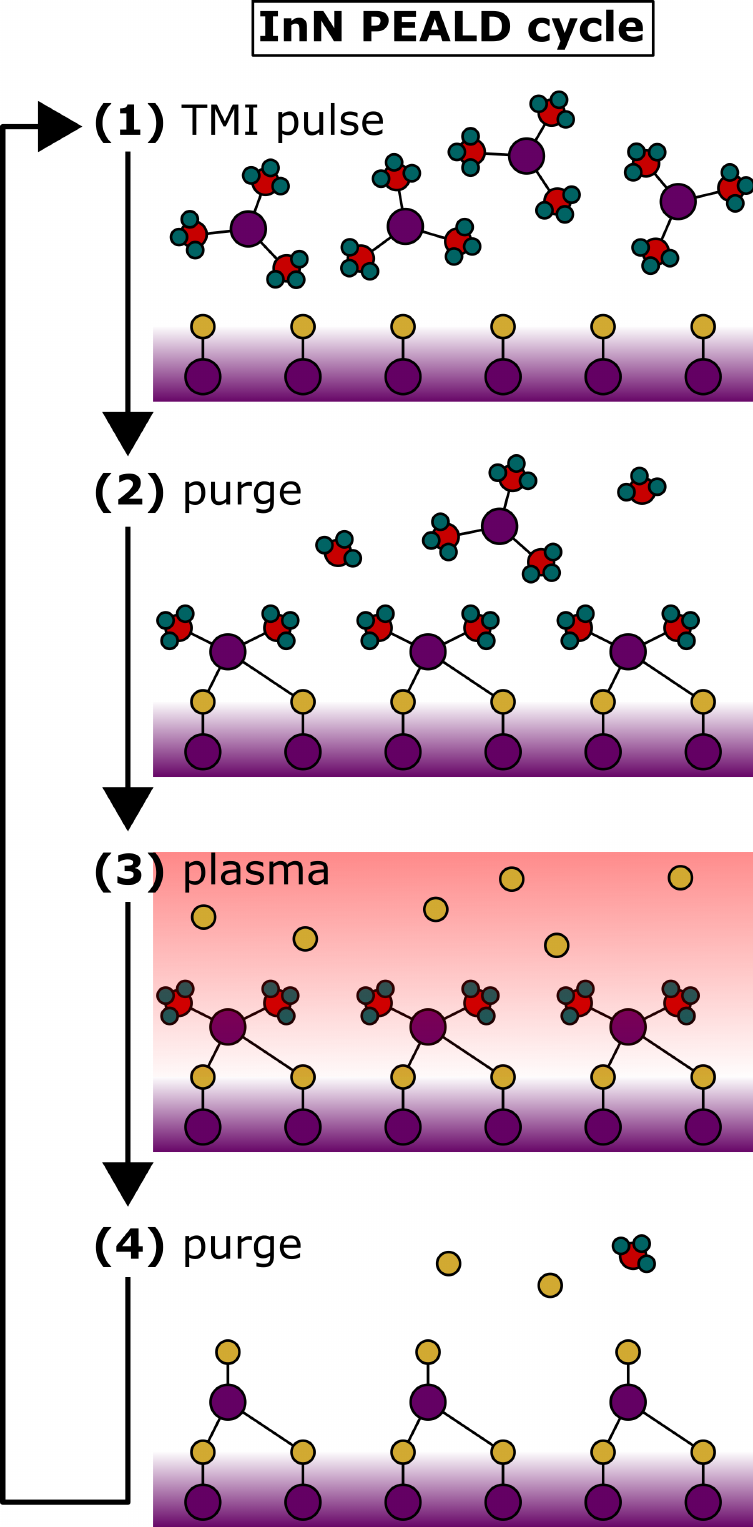}
\caption{Schematic diagram of a PE-ALD cycle. (1) TMI is pulsed into the UHV chamber where it saturates the growth surface; (2) After chemisorption to the surface, excess TMI and reaction byproducts are purged; (3) The methylindium adlayer is exposed to N$_2$/Ar plasma, reacting to form InN; (4) Plasma species and reaction byproducts are purged. }
\label{fig:PE-ALDprocess}
\end{figure}

For these experiments a partly coherent X-ray beam of 10 $\mu$m $\times$ 10 $\mu$m size impinged at a grazing-incidence angle of $\alpha_i$ = 0.38$^\circ$ during PE-ALD growth of InN on an a-plane sapphire substrate, as shown in Fig. \ref{fig:setup}.  %
The InN film was grown at 250$^\circ$C using a 200-cycle PE-ALD process.  The individual cycles consisted of four stages: 1) 60 millisecond Trimethylindium (TMI), the indium precursor, pulse; 2) 11 second purge; 3) 20 second N$_2$/Ar plasma exposure; and 4) 31 second purge. An individual PE-ALD cycle is explained schematically in Fig. \ref{fig:PE-ALDprocess}. Under the experimental conditions, it takes 8-10 PE-ALD cycles to form a monolayer of wurtzite InN, with a monolayer being half a unit cell height \cite{nepal2017real,nepal2019understanding}.

\section*{Results}\label{result_and_discussion}

\subsection*{Overview and Speckle-Averaged Kinetics}

\begin{figure}[h!]
\centering\includegraphics[width=9cm]{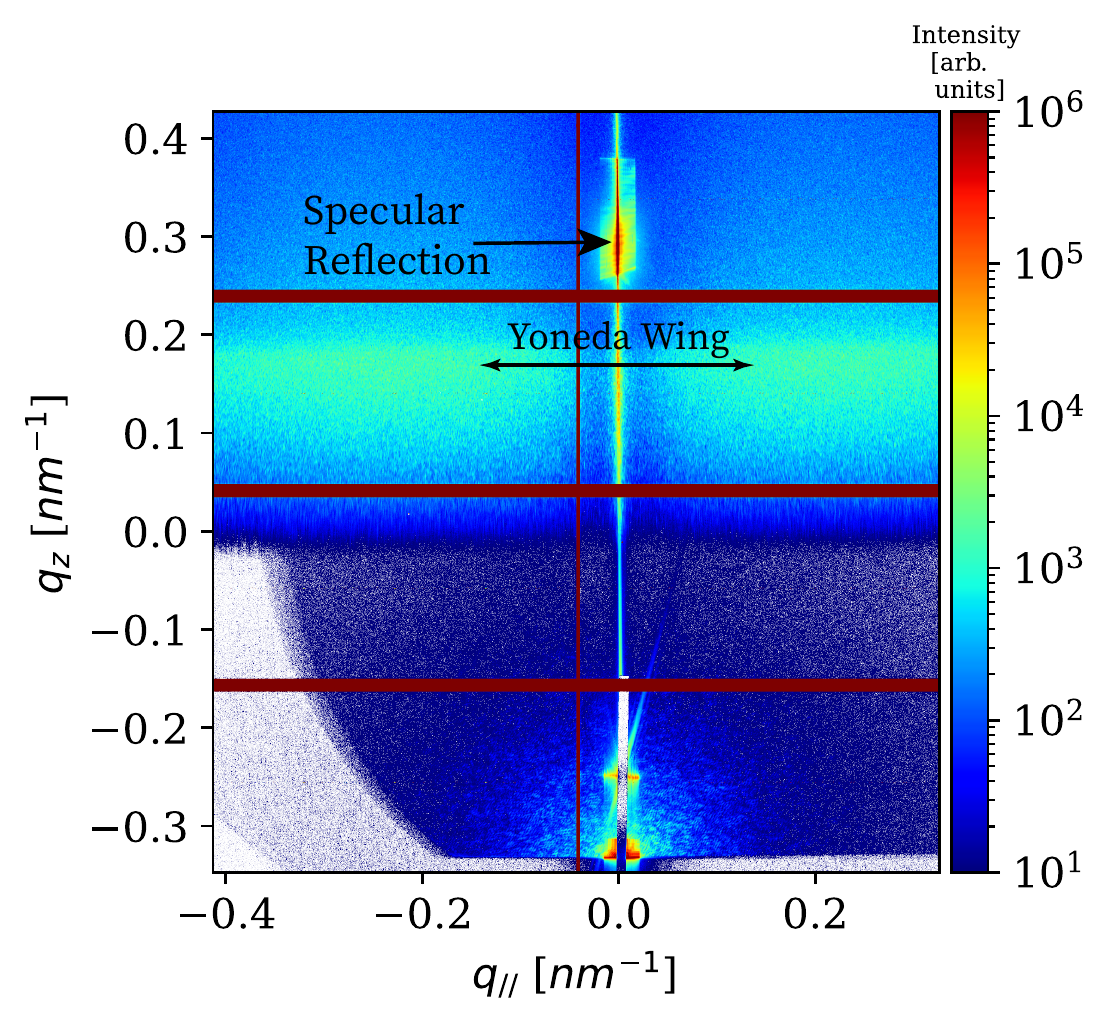}
\caption{Detector image at growth time $T$ = ($t_1$ + $t_2$)/2 = 500 s, where $T$ = 0 s is the start of first PE-ALD cycle. In this plot, the Yoneda wing is fully formed and stretches across most of the $q_{\parallel}$ range probed. Brown regions are module gaps of the detector.}
\label{fig:detectorimage}
\end{figure}

Figure \ref{fig:detectorimage} shows a typical detector image during deposition. The wave vector change during the scattering is defined as
\begin{equation}
\mathbf{q} = \mathbf{k}_f - \mathbf{k}_i
\label{eqn:def_q}
\end{equation}
$q_{\parallel}$ is the {\bf q} component in the plane of the sample while $q_z$ lies nearly perpendicular to the surface plane. %
During growth, scattering intensities increase across a broad band around $q_z = 0.1 - 0.2$ nm$^{-1}$. The stretch of these developing intensities, called the Yoneda wing, corresponds to the highly surface-sensitive scattering leaving the sample at the critical angle for total external reflection. The distinct peaks on each side of the $q_{\parallel} = 0$ line are due to the growth of spatially correlated mounds on the surface.  

\begin{figure}[h!]
\centering\includegraphics[width=9cm]{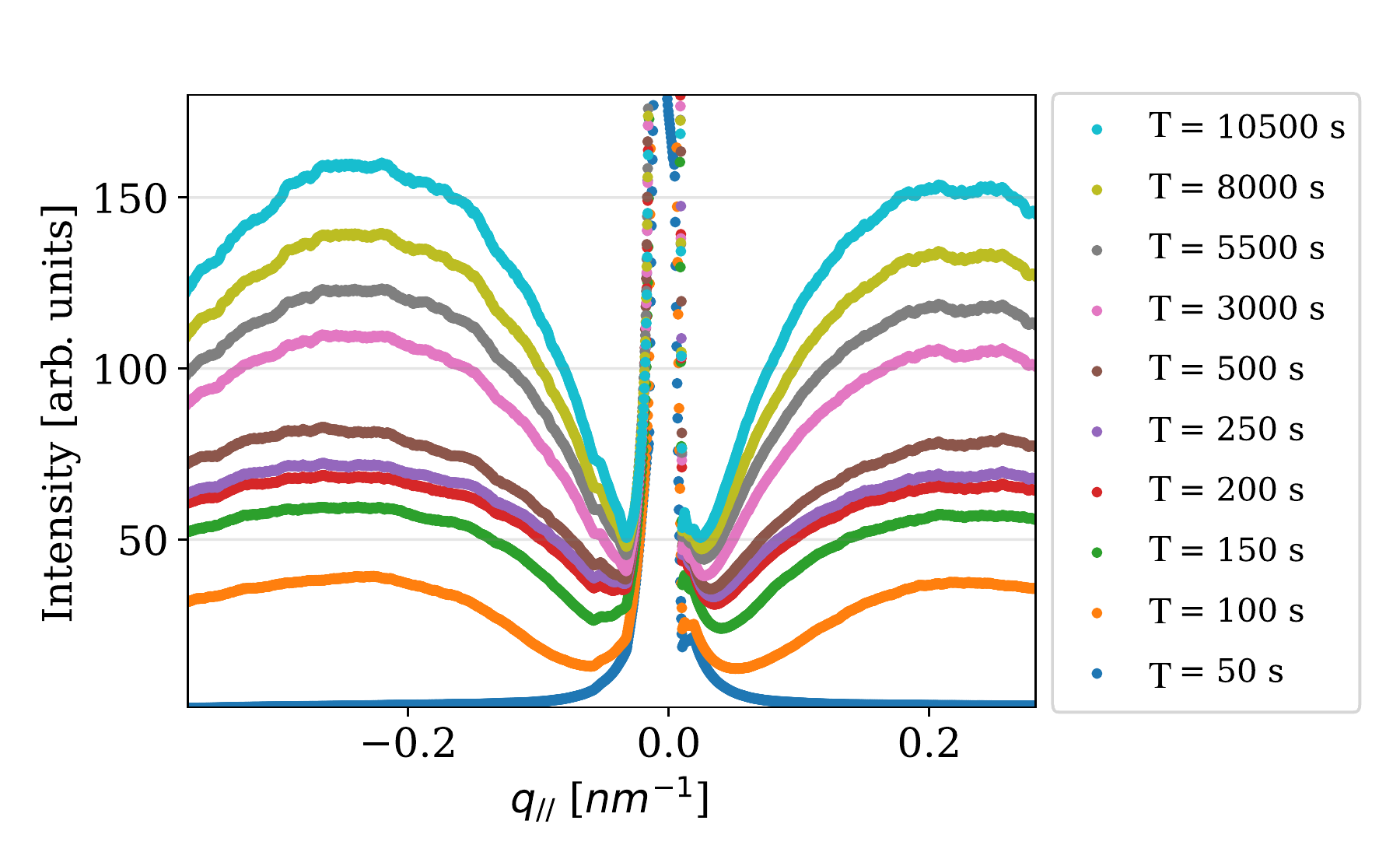}
\centering\includegraphics[width=9cm]{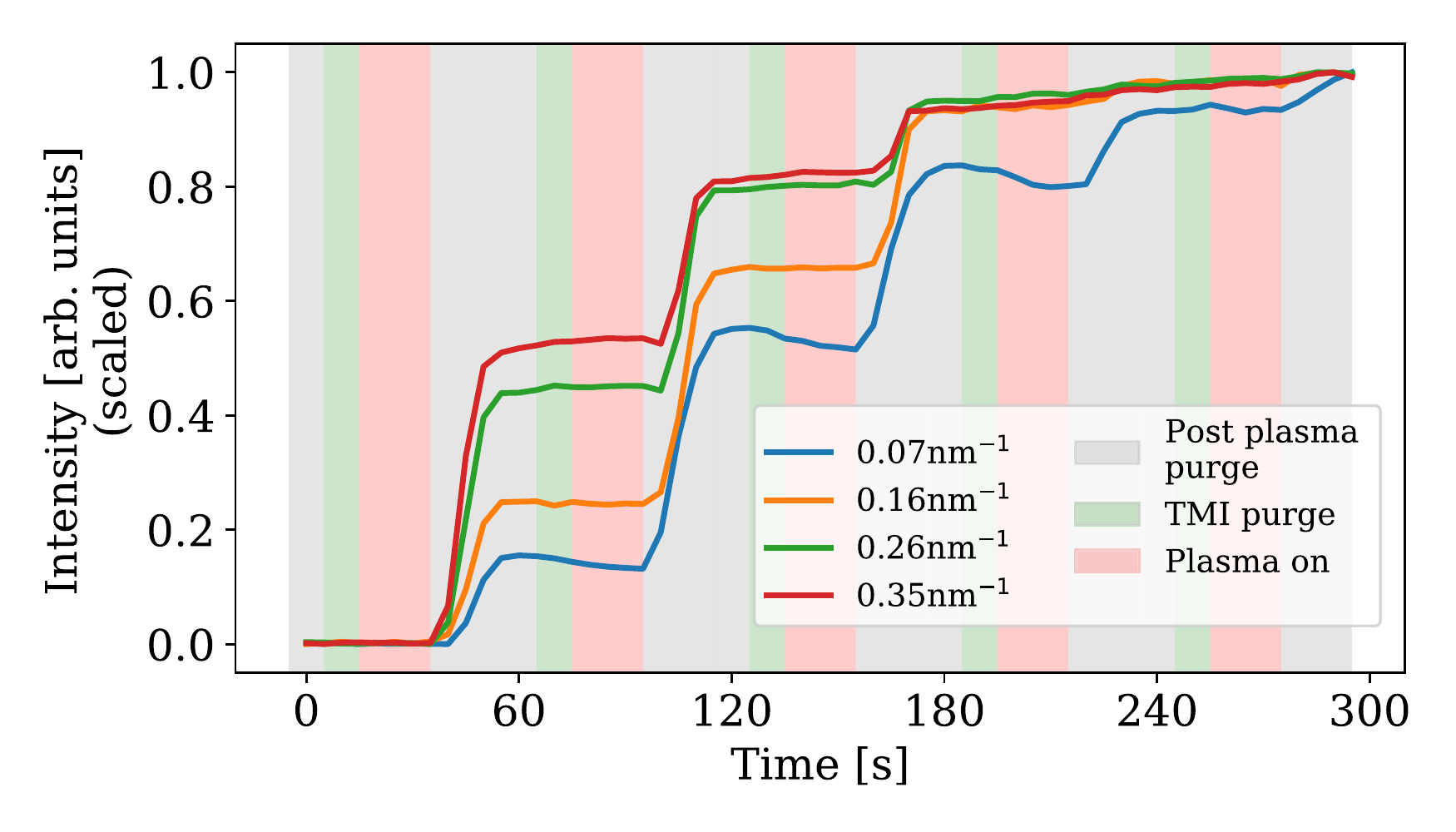}
\caption{(Top): Speckle-averaged intensity patterns at different growth times $T$ as a function of $q_{\parallel}$ at the Yoneda wing. (Bottom): Evolution of speckle-averaged intensity at different $q_{\parallel}$ as a function of time. For comparison purposes, the intensity at each wavenumber is normalized to the intensity at 300 s.}
\label{fig:intensityplot}
\end{figure}

To better understand the overall surface evolution and make correspondence with earlier studies using low-coherence X-ray scattering, we first examine the evolution of the speckle-averaged intensity along the surface-sensitive Yoneda wing as a function of $q_{\parallel}$. Since the PE-ALD-grown InN surface topography has a low Root-Mean-Square (RMS) roughness \cite{nepal2019understanding}, the speckle-averaged scattered intensity at a given $q_{\parallel}$ is approximately proportional to the square of the roughness on the corresponding length scale of 2$\pi$/$q_{\parallel}$. Resulting speckle-averaged scattering patterns at different points in the film growth are displayed in the top part of Fig. \ref{fig:intensityplot}.
Little scattering is observed at the earliest deposition time shown, 50 s, corresponding to a small fraction of monolayer deposition. But, as InN layers are formed, the scattering quickly develops a peak on each side of $q_{\parallel} = 0$ which extends beyond the edge of the detector. The peaks grow with ongoing deposition, shift slightly to lower wavenumbers, and eventually reach a stage in which the speckle-averaged intensity is changing only slowly. These observations are consistent with previous low-coherence real-time X-ray studies of the growth process \cite{nepal2019understanding,woodward2019influence}.  The appearance of correlation peaks within the earliest cycles (see Fig. \ref{fig:intensityplot}) and their gradual shifting towards lower $q_{||}$ indicates that the InN growth initially proceeds via the nucleation of correlated islands (i.e. Volmer-Weber growth) which gradually coarsen and coalesce upon sufficient lateral growth. The late stage average distance between correlated islands is $2\pi/q_{\parallel}^{peak} \approx 24$ nm.

The bottom of Fig. \ref{fig:intensityplot} shows the intensity evolution during the initial film growth for different $q_{\parallel}$.  It's seen that there is a periodic jump in intensity at deposition times of 45-50 s, 105-110 s, 165-170 s and 225-230 s.  These correspond to the time periods immediately after the plasma is turned off during the first four growth cycles.  
This timing suggests that the initial island morphology is defined during the plasma-off state of the cycle. Further information and a rationale for this behavior comes from the speckle correlation analysis and discussion below.

\subsection*{Speckle Correlation Study of Dynamics}

\subsection*{Overall Behavior}

The Two-Time Correlation Function (TTCF) $C(q_{\parallel},t_1,t_2)$ compares the exact state of the surface on a given length scale $2\pi/q_{\parallel}$ at time $t_1$ with the state at a time $t_2$:

\begin{equation}
C(q_{\parallel},t_1,t_2)= \frac{\left\langle I(q_{\parallel},t_1)I(q_{\parallel},t_2)\right\rangle }{\left\langle I(q_{\parallel},t_1)\right\rangle \left\langle I(q_{\parallel},t_2)\right\rangle} 
\label{eqn:twotime}
\end{equation}

\begin{figure}[h!]
\centering\includegraphics[width=5.9cm]{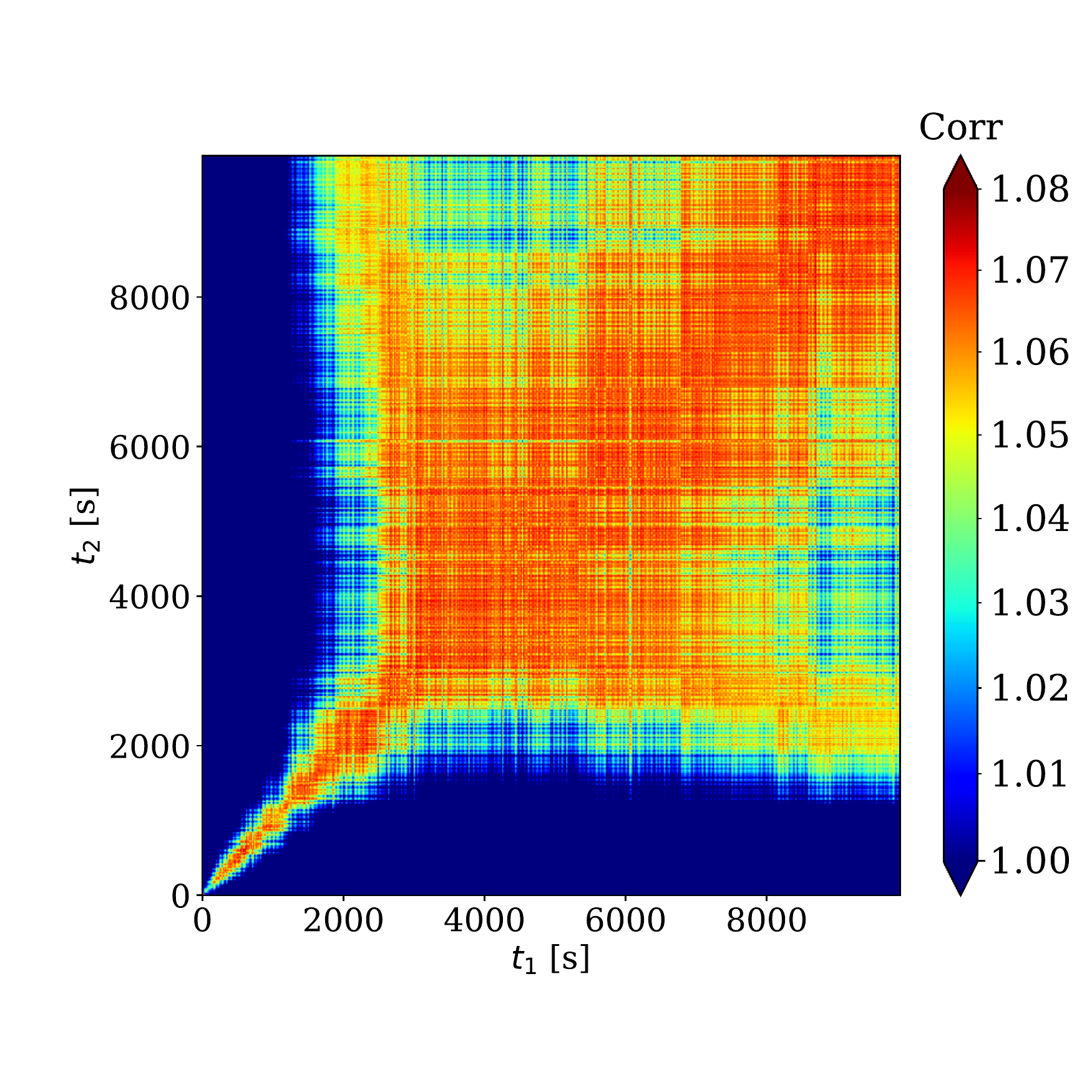}
\centering\includegraphics[width=5.9cm]{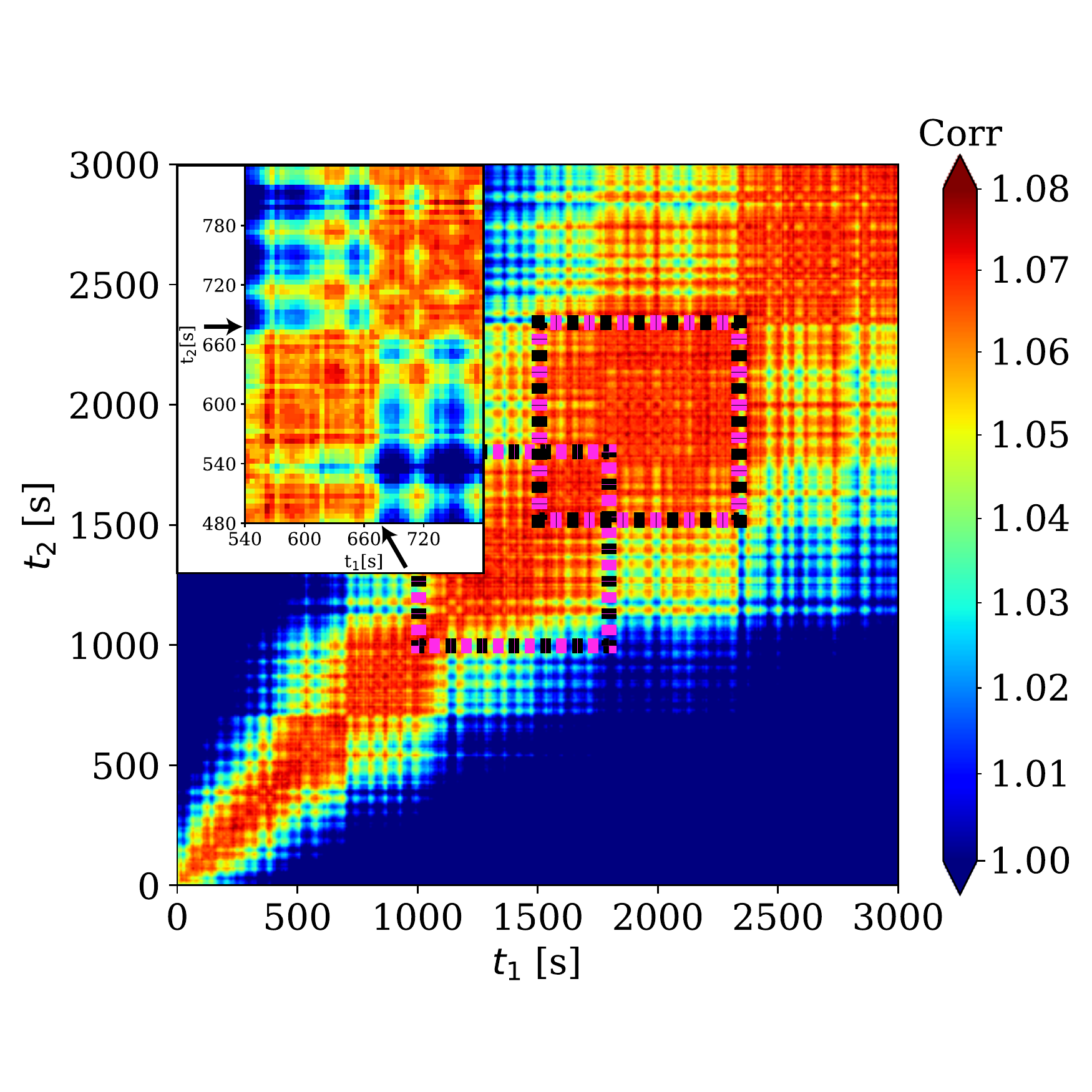}

\caption{(Left): TTCF calculated throughout the entire experiment at $q_{\parallel}$ = 0.2 nm$^{-1}$ on the Yoneda wing. Steady state is reached at t$_1$ \& t$_2$ $>$ 2000s, where the correlation is preserved throughout the rest of the TTCF. (Right): TTCF in the early time. Individual PE-ALD cycles are in 60 s $\times$ 60 s scale and appear here as bright correlation peaks (Fig. \ref{fig:TTCF_1cycle_4cycles} shows individual-cycle TTCFs). The irregular box-like multiple-cycle-covering areas of high correlation along the diagonal may be suggestive of stress relief after 5-15 PE-ALD cycles in the early time. Two such overlapping areas are annotated here with alternating black and pink dotted boxes. An area of the TTCF, covering 4 cycles (along t$_1$) by 6 cycles (along t$_2$), is zoomed in, as shown in the inset plot and the black arrows point the last time frame of TMI pulse in a cycle, where sharp correlation change is observed.}
\label{fig:TTCF}
\end{figure}

As an example, the TTCF for a region on the surface-sensitive Yoneda wing near the correlation peak is shown in the left of Fig. \ref{fig:TTCF}. A general characteristic of such correlation functions is that they are highest along the diagonal $t_1 = t_2$.   The structure of the TTCF here is very different from the parallel correlation lines observed in the TTCF when layered 2D growth plays the dominant role \cite{headrick2019coherent,ju2019coherent}. For the present PE-ALD growth process, the left plot of Fig. \ref{fig:TTCF} shows that, off the diagonal, i.e. $t_1 \neq t_2$, correlations of the growth surface topography are initially very short-lived but gradually extend to longer time differences $\Delta t = \vert t_2 - t_1  \vert$ as the 3D growth proceeds. Throughout the growth process, a repetitive structure is observed with the 60 s period of the PE-ALD growth process.  This behavior is observed at all $q_{\parallel}$ and will be discussed later; we first focus on more general characteristics of the TTCF.%

During the early stages of growth, the TTCF shown in the right graph of Fig. \ref{fig:TTCF} exhibits an irregular box-like pattern indicating periods of relatively little change to the surface morphology, consistent with generally conformal growth over established 3D features, separated by sudden changes to the scattering pattern.  %
This pattern suggests the existence of sudden nanoscale structural changes on the surface.  Similar behavior was observed during an XPCS study of a martensitic transition in cobalt and was associated with sudden stress relief events \cite{sanborn2011direct}. The time between these events appears to increase with film thickness during the early stages of growth. Detailed examination of the timing of these sudden events (inset in right graph of Fig. \ref{fig:TTCF}) shows that they occur primarily during the purge following the precursor exposure.  This is discussed further below, in the Discussion section.

Around the deposition time of $T$ = 2000 s, corresponding to an average film thickness of slightly over 4 layers, the persistence time of correlations suddenly grows and becomes comparable to the length of the experiment.  This shows the formation of a relatively long-lived microstructure, i.e. the growth has become highly conformal on the nanoscale. Within the context of the 3D growth morphology, the mounded configuration is very stable after this point.  Nepal {\it et al.} \cite{nepal2017real} suggested that around 2-unit cell thickness, corresponding to 4 layers, the substrate surface is more than 50\% covered by the film and the mound structure stabilizes.  These experiments provide direct evidence for the latter part of that supposition. 

\subsection*{Cyclic Behavior}

\begin{figure}[h!]

\centering\includegraphics[width=9.4cm]{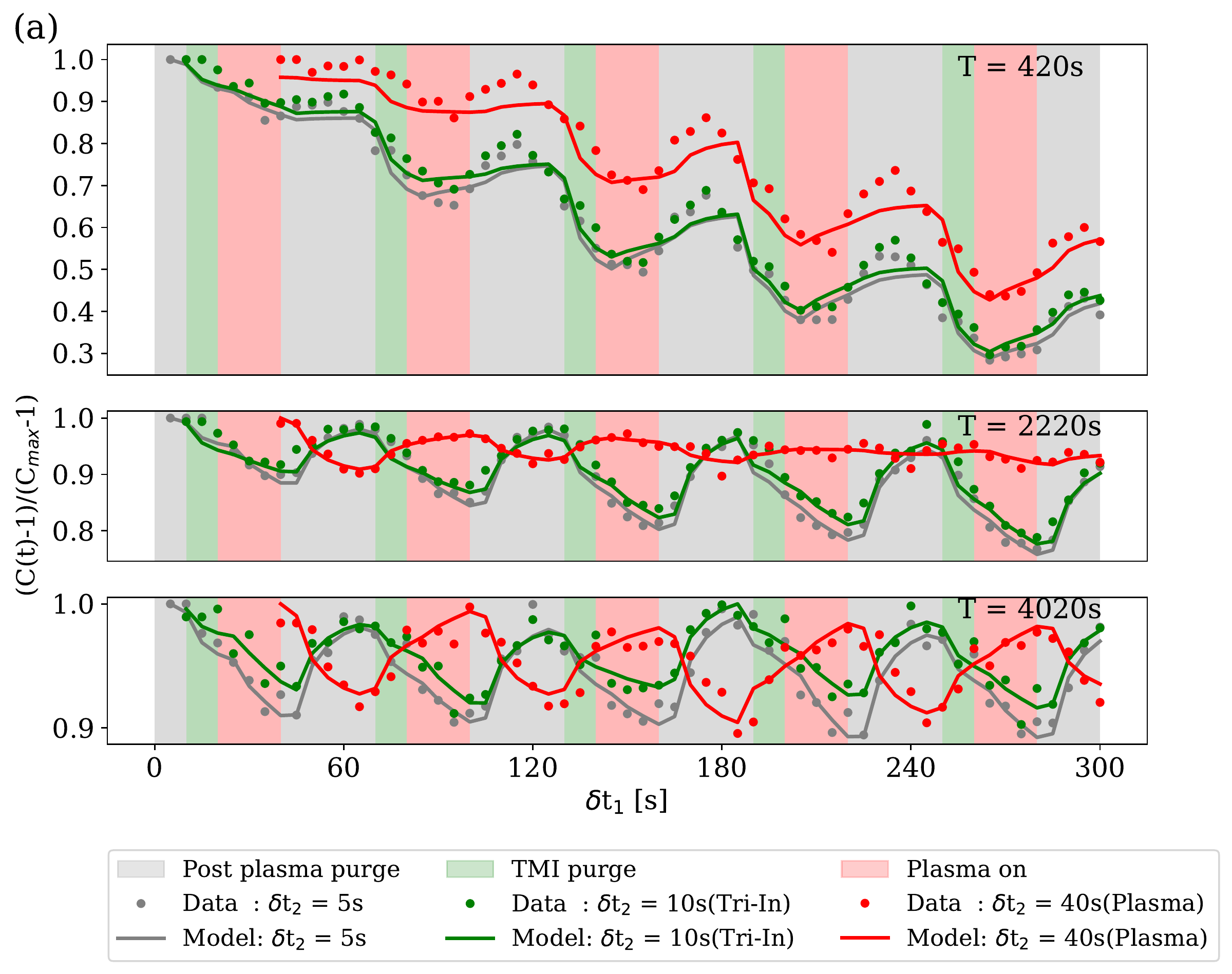}

\centering\includegraphics[width=9.7cm]{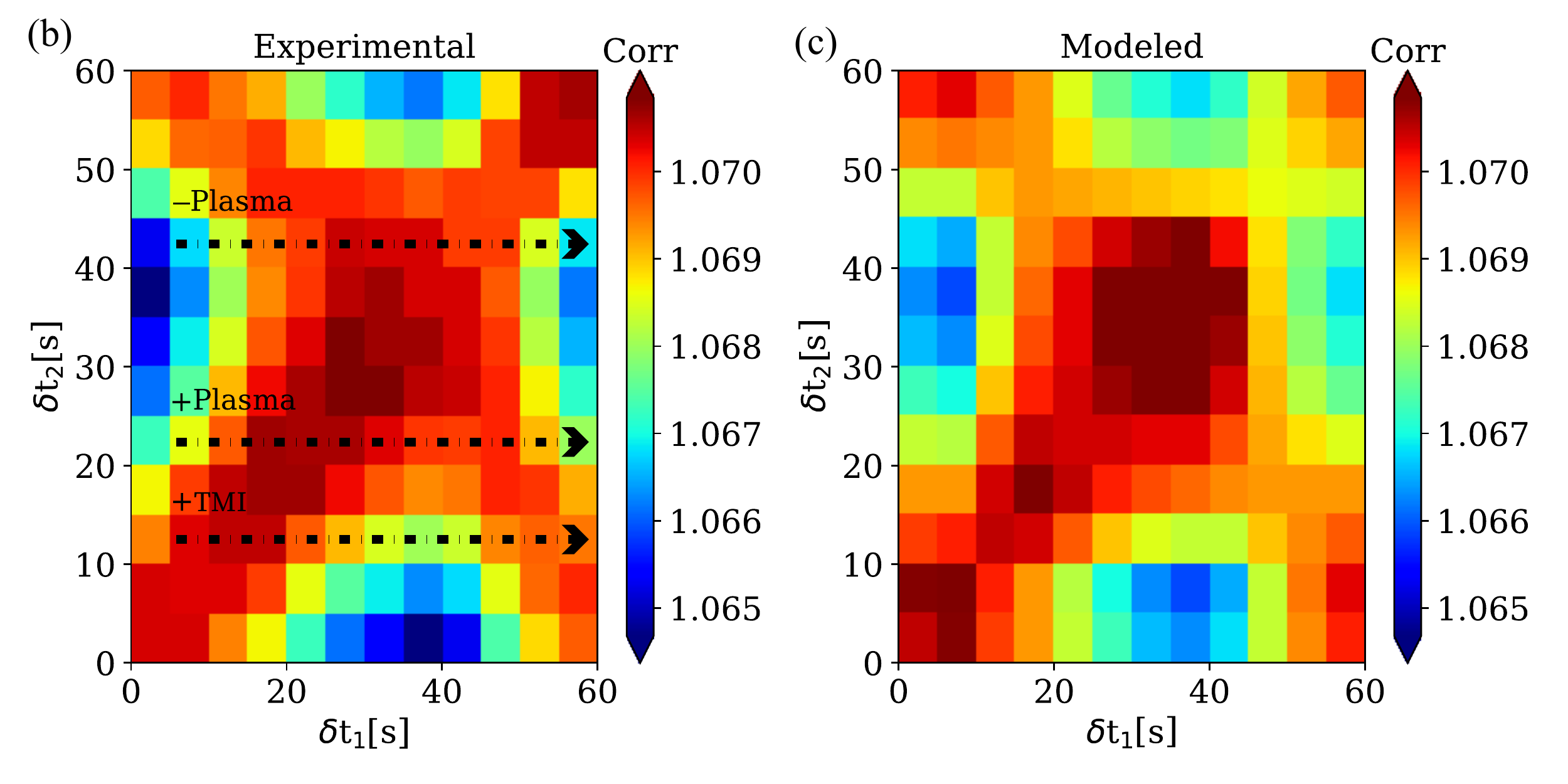} 

\centering\includegraphics[width=9.4cm]{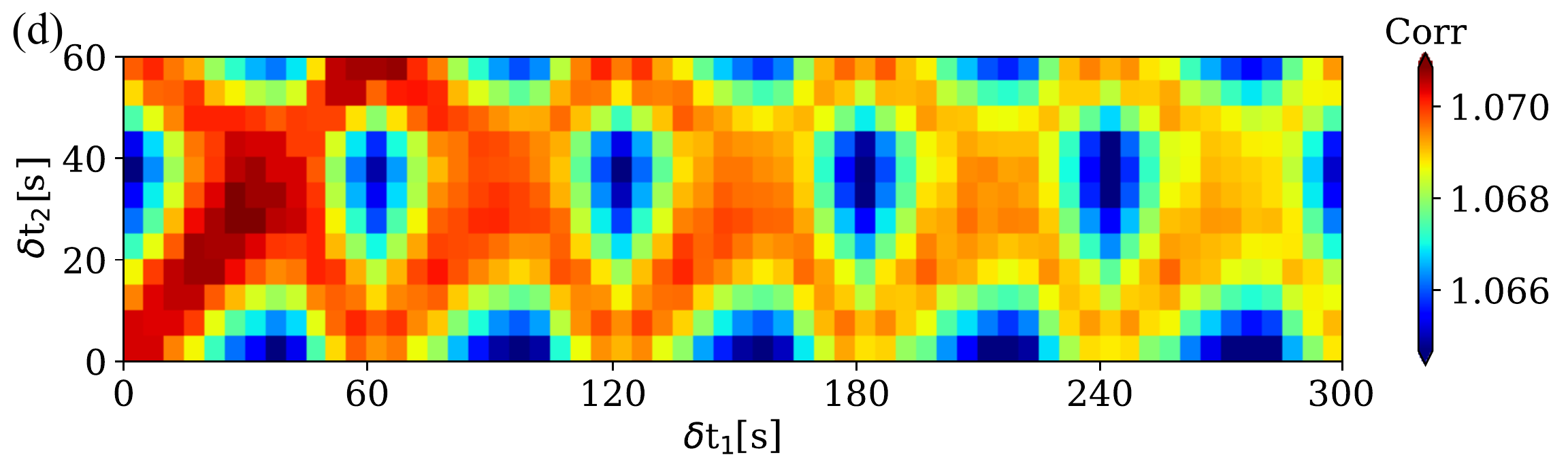} 
\caption{(a): Line cuts and model fits at different average growth times $T$ as a function of $\delta t_1$ for different $\delta t_2$. (b): Experimental TTCF of a single cycle at $T$ = 4020 s, along the $t_1 = t_2$ diagonal line averaged over 21 PE-ALD cycles along $t_2$ (c): Calculated TTCF from the model of Eq. \ref{eqn:specycle}. (d): Experimental TTCF from (b) with sequential cycles.}
\label{fig:TTCF_1cycle_4cycles}
\end{figure}

As noted above, beginning at the earliest times of the growth process there is a fine mesh structure in the TTCF associated with the 60 s PE-ALD growth cycle. To explore this behavior in more detail and examine how it evolves during growth, horizontal line cuts of the TTCF are made.  The scattered X-ray signal is relatively modest, and is particularly weak during the initial film growth stages.  To overcome this issue while preserving dynamics information over longer time scales, correlation evolution at a given average growth time $T$ = ($t_1$ + $t_2$)/2 was computed by averaging the TTCF over 15 growth cycles in $t_2$ around the average growth time $T$ (see description in Methods section below). Resulting averaged correlation functions are shown in  \ref{fig:TTCF_1cycle_4cycles}(a); here $\delta t_1$ and $\delta t_2$ denote the elapsed time from the beginning of a cycle specified by the overall average growth time $T$. For a given average time $T$ in the overall film growth process, the figure displays how much the surface at a later relative time $\delta t_1$ duplicates the surface at an earlier relative time $\delta t_2$.  The duplication is a maximum when $\delta t_1 = \delta t_2$, i.e. when a given state of the surface morphology is compared to itself, but our primary focus is on $\delta t_1 > \delta t_2$.  The zero of relative time $\delta t$ here is chosen to be during the post-plasma purge cycle.  The timing of events on this scale is: 1) end of post-plasma purge from $\delta t$ = 0 - 10 s; 2) 60 ms TMI precursor pulse; 3) purge from $\delta t$ = 10 - 21 s; 4) N$_2$ gas turns on and plasma then turns on and runs from $\delta t$ = 22 - 42 s; 5) beginning of post-plasma purge from $\delta t$ = 42 - 60 s, which leads back into step 1.
In examining these correlations, one must remember that each frame corresponds to a 5 s detector integration.

The top part of Fig. \ref{fig:TTCF_1cycle_4cycles}(a) shows the evolution of correlations at the relatively early average deposition time $T$ = 420 s. There is both a cyclic structure associated with the growth cycle and a decrease in overall correlations from one cycle to the next. This is independent of the initial state of the surface, occurring whether the initial comparison state is during the post-plasma purge ($\delta t_2$ = 5 s), after the precursor deposition ($\delta t_2$ = 10 s) or during the plasma exposure ($\delta t_2$ = 40 s).  The decrease in correlation, reflecting surface morphology evolving away from the original state, occurs almost entirely during the latter part of the post-plasma purge and the subsequent precursor deposition and purge.  During the plasma exposure and early part of the post-plasma purge, the structure moves back toward the initial state.

As growth proceeds and a continuous film is formed, the middle part of Fig. \ref{fig:TTCF_1cycle_4cycles}(a), for average deposition time $T$ = 2220 s, shows that the surface responds differently than it had earlier in the growth process.  Now there is significantly less change in overall surface morphology with time, marking a more highly conformal growth. But there is also significant difference in behavior depending upon whether the initial comparison point is the post-plasma purge ($\delta t_2$ = 5 s) state or precursor-deposited ($\delta t_2$ = 10 s) state versus the plasma-exposed ($\delta t_2$ = 40 s) state. The evolution of the surface structure away from the initial post-plasma purge and precursor-deposited states follows the behavior observed earlier.  But now the decrease in correlations away from an initial plasma-exposed state is much more muted. As subsequent growth cycles proceed, the surface continues to strongly duplicate the initial surface in the plasma-on state and decreasingly duplicate the surface as it was in the post-plasma purge and precursor-deposited states. As a result, the curve for the plasma-exposed $\delta t_2$ = 40 s state is now out of phase with the curves for the post-plasma purge ($\delta t_2$ = 5 s) and precursor-deposited ($\delta t_2$ = 10 s) states.  This trend is amplified at later growth times in the bottom part of Fig. \ref{fig:TTCF_1cycle_4cycles}(a), for average deposition time $T$ = 4020 s.  Correlations now remain high overall through subsequent growth cycles, indicative of a strong overall conformality, but there is a now a clear back-and-forth evolution of modest size between the plasma-on and plasma-off states through the course of the growth cycle.

To further explore this periodic evolution at late growth times within an individual cycle, as well as between neighboring cycles, data from 21 PE-ALD cycles along the $t_1 = t_2$ diagonal of TTCF centering $T$ = 4020 s 
are averaged along $t_2$ for better statistics to produce a single-cycle TTCF shown in Fig. \ref{fig:TTCF_1cycle_4cycles}(b). Although Figs. \ref{fig:TTCF} and \ref{fig:TTCF_1cycle_4cycles} are calculated for $q_{\parallel}$ = 0.2 nm$^{-1}$, similar behaviors are seen at all other wavenumbers within the detector range. We see that within a cycle, the TTCF exhibits an ``X" shape, immediately showing that there is significant evolution back and forth between surface structures when plasma is turned on and then turned off. This behavior carries on through sequential PE-ALD cycles as shown in Fig. \ref{fig:TTCF_1cycle_4cycles}(d).  %

Consider first individual rows going left to right through the TTCF of Fig. \ref{fig:TTCF_1cycle_4cycles}(b) at constant $\delta t_2$.  The bottom row corresponds to relative $\delta t_2$ = 5 s within the start of a growth cycle.  Correlations initially remain high, but then decrease with increasing $\delta t_1$ during the aftermath of the precursor deposition and as the plasma nitridation process then proceeds.  After the plasma is turned off at approximately $\delta t_1$ = 42 s, correlations again grow during the post-plasma purge to largely recover the original ``plasma-off" correlations present at the beginning of the cycle. Thus, during a growth cycle, the surface morphology changes away from the initial purge state, reaches a maximum of difference at the time the plasma is turned off, and then returns to a state rather similar to the original state.   

Examination of the row corresponding to $\delta t_2$ = 40 s, i.e. the last frame with the plasma fully on, reinforces this picture. The initial state with $\delta t_1$ = 5 or 10 s (during the post-plasma purge) has a maximal difference with the morphology created by the plasma.  There is only a small change of correlations when the TMI pulse first arrives, but the correlations continue evolving afterward.  When the plasma turns on ($\delta t_1$ = 22 s), the structure evolves continuously toward the final plasma-on state.  After the plasma is turned off, the structure relaxes continuously back toward the plasma-off structure.

Examination of the row corresponding to $\delta t_2$ = 25 s provides additional information.  Notably, correlations reach a maximum after the start of the plasma exposure, then decrease as the plasma process continues. After the plasma is turned off ($\delta t_1$ = 42 s), and the post-plasma purge occurs, correlations initially increase again, reaching a maximum in the range of $\delta t_1$ = 50 s, but then decrease as the purge continues. This double-peak behavior at this $\delta t_2$ suggests that the structures through which the surface passes during plasma nitridation are the same or similar to those through which it passes after the plasma is turned off.  

We now turn to modeling the results before further discussing their implications.

\section*{Modeling}\label{sec13}

The behavior of the TTCF within a PE-ALD cycle can be understood using models incorporating transitions between different surface states.  For investigation of the late time growth as exhibited in \ref{fig:TTCF_1cycle_4cycles}(b), we explore a model with four states: 1) asymptotic plasma-on (``on"); 2) asymptotic plasma-off (``off"); 3) precursor-deposited (``pre"); and 4) precursor-relaxed (``rel"), i.e. the state toward which the surface is heading in the period following precursor deposition.  The speckle patterns of the four states are designated $I_{on}({\bf q}), I_{off}({\bf q})$, $I_{pre}({\bf q})$ and $I_{rel}({\bf q})$ respectively.  %

In calculating the TTCF from the model, the variable parameters are the overall speckle contrast $\beta$, the correlations between the different states $C_{\alpha-\xi} = \langle I_{\alpha}({\bf q})I_{\xi}({\bf q}) \rangle / I_{ave}^2$, and the three relaxation times for changing surface structure $\tau_{pre}$, $\tau_{on}$ and $\tau_{off}$.  These are the characteristic times for the evolution of the surface after precursor deposition, after the plasma is turned on, and after the plasma is turned off, respectively.

Figure \ref{fig:TTCF_1cycle_4cycles}(c) shows a model fit that is in semi-quantitative agreement with the experimental data of Figure \ref{fig:TTCF_1cycle_4cycles}(b).  Tables \ref{tab:C_model} and \ref{tab:tau_alpha} list the relative correlation values ($(C_{\alpha-\xi}-1)/\beta$ and correlation times used in the model. It's seen that the relative correlation of the precursor-deposited state with the plasma-off state is essentially 100\% and the correlations of those two states with others are also essentially identical.  Thus, there is a high degree of conformality when the precursor is deposited - in accord with the simple picture that the precursor molecules bind at all possible sites within the constraints of steric hindrances.

However, after deposition we see that the surface starts evolving to the ``relaxed" state which has relative correlations close (0.97) to that of the plasma-on state.  Thus, even before the plasma is turned on, the surface morphology is evolving toward a state that is somewhat similar to that ultimately reached during the plasma exposure. 

\begin{table*}
\centering
\begin{tabular}{|*{5}{c|}}\hline
    State ($\alpha$/$\xi$) & on & off & pre & rel \\ \hline
    on & 1.00 & 0.87 & 0.87 & 0.97 \\ \hline
    off & 0.87 & 1.00 & 1.00 & 0.92 \\ \hline
    pre & 0.87 & 1.00 & 1.00 & 0.92 \\ \hline
    rel & 0.97 & 0.92 & 0.92 & 1.00 \\ \hline
\end{tabular}

\caption{The relative correlations $(C_{\alpha-\xi}-1)/\beta$ calculated from the four-state model. The table is symmetric about the diagonal line because $C_{\alpha-\xi} = C_{\xi-\alpha}$. For the model, $\beta$ is fit to be 0.073.}
\label{tab:C_model}
\end{table*}

\begin{table*}
\centering
\begin{tabular}{|*{4}{c|}}\hline
    $\alpha$ & on & off & pre \\ \hline
    $\tau_\alpha$(s) & 11.1 & 12.1 & 7.1 \\ \hline
\end{tabular}

\caption{The $\tau$ parameters used for the model calculation shown in Fig. \ref{fig:TTCF_1cycle_4cycles}(b).}
\label{tab:tau_alpha}
\end{table*}

The model can be extended to fit the evolution of correlations at earlier growth times $T$ shown in Fig. \ref{fig:TTCF_1cycle_4cycles}(a). For simplicity and building upon the late time observations, here we neglect any difference in surface morphology between the plasma-off state and the precursor-deposited state.  At each new point in the cycle, evolution of the precursor-deposited surface, plasma turned on and plasma turned off, there is assumed to be an exponential relaxation toward a new state.  Because the structural correlations decay with advancing growth time during the early and mid-periods of the growth, the correlations between the initial state and each subsequent growth stage are allowed to vary in the fit.  Results are shown as the solid lines in Fig. \ref{fig:TTCF_1cycle_4cycles}(a) and reproduce the data trends (points) reasonably well.

\section*{Discussion}\label{sec15}

While state-of-the-art real-time studies using conventional approaches have identified structural evolution during precursor deposition and plasma exposure \cite{nepal2017real,nepal2019understanding,woodward2022influence,ilhom2020elucidating}, the uniquely detailed view provided by coherent X-ray scattering suggests a complex picture.  The plasma does not simply freeze in a structure which is then built upon in subsequent cycles.  The XPCS results show that there are significant changes in nanoscale film surface structure occurring in all parts of the cycle, including purge periods following the precursor saturation and plasma exposure of the surface.  Moreover, the results collectively document how the nanoscale dynamics evolve as the film growth proceeds throughout the process.

\begin{figure}[h!]
\centering\includegraphics[width=11cm]{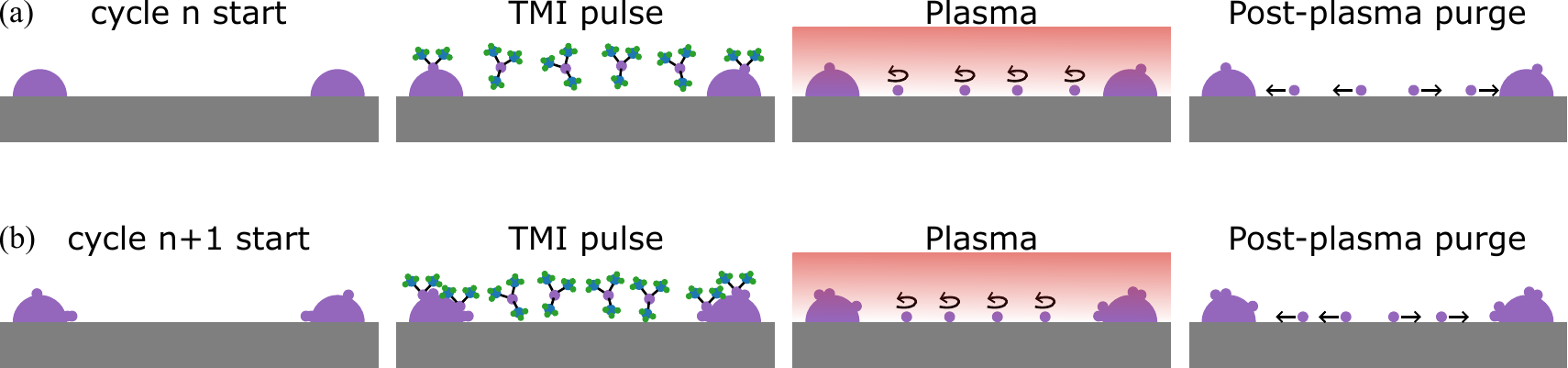}

\caption{Illustration depicting the proposed dynamics during the early stages of growth for arbitrary (a) cycle n and (b) cycle n+1.}
\label{fig:proposed_dynamics_cartoon}
\end{figure}

During the initial growth cycles on the sapphire substrate, the speckle-averaged intensity (bottom plot of Fig. \ref{fig:intensityplot}) only increases during the initial part of the post-plasma purge, which indicates that it is during this part of the cycle that the mound structure grows. This in turn implies that the adatom species migrate farthest on the bare sapphire during the initial part of the post-plasma purge, allowing them to coarsen existing InN islands or coalesce into indium nanodroplets.  This contrasts with an earlier low-coherence real-time X-ray study of PE-ALD growth of InN on GaN which found that initial correlated island growth was associated with the plasma nitridation part of the growth cycle \cite{woodward2022influence}.  The difference could be due to the different substrates - adatom diffusion on the sapphire surface is generally slower than on GaN \cite{gao2013epitaxial}. The In diffusion length is also limited when there is a background of active nitrogen species.

More detailed information about the continued early growth period comes from the TTCF at $T$ = 420 s when there is sufficient intensity to measure the TTCF but regions of bare substrate surface remain.  Here we see that correlations with the initial state are preserved or grow during the earliest part of the post-plasma purge. However, during the latter part of the post-plasma purge (top graph in \ref{fig:TTCF_1cycle_4cycles}(a)), correlations then decay, and hence the surface morphology evolves most away from its earlier state. 

A cartoon illustrating a proposed early time dynamics consistent with the early-stage growth results is shown in Fig. \ref{fig:proposed_dynamics_cartoon}. At the beginning of a cycle the surface consists of a distribution of separated islands on the substrate. TMI is then pulsed into the chamber, saturating the entire surface. Due to the constraints of steric hindrance, adsorbed species occupy only a fraction of available lattice sites, and there is a degree of randomness to which sites are occupied. During plasma exposure, the ligands are stripped from the methylindium species and the indium reacts with active nitrogen species to form InN. However, the new InN nuclei on the sapphire surface are inherently unstable due to their small size. Thus, throughout the plasma exposure, adatoms continuously react with atomic N to form unstable InN nuclei, decompose, and diffuse short distances. When the plasma is turned off, In adatoms can diffuse greater distances to join existing InN islands or coalesce with other adatoms. Initially, existing islands are reinforced in the plasma-off process while the sapphire surface is gradually cleared due to a combination of adatom diffusion towards islands and desorption of reaction byproducts.  Both lead to increasing correlation with the structure existing at the beginning of the cycle.

The box-like structures in the TTCF during these early stages of growth are consistent with numerous overlapping stress-relief events, presumably due to sudden stress relaxation events associated with the mismatch film stress and island coalescence. It can be expected that the InN film initially contains a nonuniform distribution of compressive and tensile stresses due to lattice mismatch with the a-plane sapphire and ongoing island coalescence, respectively, which evolve with thickness \cite{lee2002synchrotron,jain2004evolution}. In addition, for Volmer-Weber growth of InN on lattice mismatched substrates, stress relaxation due to the formation of misfit dislocations at island/substrate interfaces has been observed to occur after initial island formation with continued growth \cite{ivaldi2011influence} rather than immediately upon nucleation. From this viewpoint, then, the irregular box-like pattern and general temporal evolution of the TTCF are consistent with the presence of a spectrum of sudden stress relief events in the growing film. As noted above, these events occur preferentially during the purge following the precursor exposure within a particular growth cycle.  We hypothesize that these are caused by changes in the InN surface structure associated with the gradual chemisorption of physisorbed TMI molecules.  While the time for complete chemisorption of physisorbed species is dependent on the reactor geometry, process environment, and surface chemistry, a study of InN PE-ALD by Hsu {\it et al.} \cite{hsu2021on} in which NH$_3$ plasma was used as the reactant found that a purge of several seconds was required.

During the early stages, the time period over which the nanoscale surface structure remains correlated gradually increases as growth proceeds as seen in the increasing width of the central ridge of the left TTCF in Fig. \ref{fig:TTCF}. However, when the deposition reaches approximately 4 layers, the nanoscale surface morphology rather suddenly becomes long-lived, with conformal correlation time spanning the period of the experiment \textemdash thousands of seconds.  An earlier study with real-time low-coherence X-ray scattering suggested that at this point the film becomes continuous and the mound structure stabilizes \cite{nepal2017real}.  However, low-coherence studies can only show that the {\it average} structure has reached a steady state evolution.  It requires coherent scattering experiments, such as used in XPCS, to monitor the evolution of the specific morphology and demonstrate the presence (or absence) of nanoscale conformal growth. %

While the TTCF is not completely uniform after 2000 s, it's no longer dominated by the irregular box-like pattern observed at earlier growth times.  This is also consistent with the situation during Volmer-Weber growth of metal films which can exhibit stress-relieving slip at the substrate/film interface during early growth stages but not after islands coalesce \cite{seel2002stress}. It's unclear whether the larger scale nonuniformity in the TTCF is due to actual variation in surface structure, perhaps associated with variations in monolayer completion dynamics arising from sub-monolayer additions of new adatoms per cycle, or whether instead it's due to difficulties in normalization of the TTCF. 

As growth proceeds beyond the point of forming a continuous film, there is now a more modest variation in surface structure from one growth cycle to the next as seen in the middle graph of Fig. \ref{fig:TTCF_1cycle_4cycles}(a). The timing of the decreasing minima with passing $\delta t_1$ for the $\delta t_2$ = 5 s and $\delta t_2$ = 10 s curves show that the overall evolution of the surface morphology is being driven by the plasma exposure. Moreover, during subsequent growth cycles, structure associated with the plasma-on state ($\delta t_2$ = 40 s) remains significantly more correlated than does structure associated with other reference states. Thus, it is again the plasma exposed state defining the future evolution of the film surface structure. Although the plasma does not simply freeze in a structure which is then built upon in subsequent cycles, on the time scale of several successive cycles, it’s the plasma-on state which retains correlations in moving from one cycle to the next.

As growth continues (bottom graph of Fig. \ref{fig:TTCF_1cycle_4cycles}(a)), the TTCF becomes increasingly dominated by a morphology that alternates between the plasma-on and the plasma-off states, with the initial plasma-on curve $\delta t_2$ = 40 s curve now being exactly out of phase with the initial plasma-off curve $\delta t_2$ = 5 s. The timing of the oscillations within the growth cycle suggests that the alternation is between 1) an active surface state (precursor exposure, purge and plasma exposure) covered with adsorbed precursor molecules and their fragments as In and N are incorporated into the existing lattice and 2) a bare surface state (post-plasma purge) toward which the morphology returns as unincorporated species leave and an uncovered surface conformal with the previous uncovered state is exposed.  As the sapphire substrate is now fully covered by InN, significant differences in the surface dynamics compared to those of the early growth stage are to be expected.  The combination of comparatively greater adatom diffusion, thermodynamic stability of topographical features, and uniformity of surface chemical potential after complete coverage of the sapphire substrate promotes a more conformal growth mode in which the morphological evolution primarily occurs during plasma exposure, as has been reported for InN PE-ALD on GaN \cite{woodward2022influence}. As for the post-plasma purge, a surface reconstruction may be responsible for its gradually transitioning surface state, as multiple evidences of surface reconstructions have been observed {\it in situ} during the growth of InN by molecular beam epitaxy \cite{himmerlich2009pambe}.

While traditional experiments do not show sub-cycle kinetics beyond the initial growth, the XPCS experiments here have allowed us to quantitatively explore the time scales on which the nanoscale surface morphology responds to each step in a growth cycle.  Detailed fitting (Fig. \ref{fig:TTCF_1cycle_4cycles}(b); Table \ref{tab:C_model}) shows that the precursor exposure itself does not significantly change the nanoscale surface structure but that, after exposure, the surface begins evolving significantly even before the plasma is turned on.  This is consistent with Muneshwar and Cadien's kinetic model of ALD surface reactions in which the precursor pulse results in a physisorbed adlayer which is then subject to various surface processes such as diffusion, desorption, and chemisorption, each with finite kinetic rates \cite{muneshwar2018surface}.  The characteristic time scales for nanoscale surface relaxation in response to the other processing steps (Table \ref{tab:tau_alpha}) are on the order of 10 s.  It's often tacitly assumed that each sub-cycle component approaches a steady state.  However here we see that there is continuing morphological evolution throughout all other components of the cycle without a steady state being reached before the next sub-cycle step begins (cf. Fig. \ref{fig:TTCF_1cycle_4cycles}(a)).  This is particularly noticeable in the case of the post-plasma purge process, for which the precursor exposure and post-exposure evolution significantly disrupt the trajectory of correlations. It’s also apparent in the interrupted trajectory of the precursor-exposed state when the plasma is turned on.

In the future, relaxation time information from XPCS could be especially useful for complicated growth processes which include a greater number of disparate stages per cycle.  For example, in the digital growth of alloys using an A-B-C-D (precursor 1, plasma 1, precursor 2, plasma 2) process rather than A-B (precursor 1, plasma 1) process, it would be helpful to determine how the long-term evolution is influenced by these individual stages.  Growth processes including a periodic atomic layer etch step to smoothen surface topography or maintain selectivity in area selective deposition would also benefit from such relaxation time information.  Furthermore, while in the classic ALD model a thermodynamically stable surface is established with each reaction step, there is also the possibility of utilizing thermodynamically unstable but kinetically stable chemistries provided that decomposition of the adlayer is sufficiently delayed \cite{pedersen2016time}.  For such an approach to ALD, access to relaxation time information would be helpful for process optimization, due to the significantly increased complexity of the role of time.

Though the present work has demonstrated how new knowledge could be extracted from the real-time study of pulsed processes such as PE-ALD, it has been restricted by the modest coherent flux available.  This has limited the time resolution and also created the need to average over many cycles to get reasonable statistics for the TTCF within a cycle.  The continued development of accelerator-based X-ray sources with higher coherent brightness will further the application of these approaches to the wide range of epitaxial thin film growth processes.

\section*{Methods}\label{sec14}
\subsection*{Experiment}
The experiment was performed at the Coherent Hard X-ray (CHX) Beamline 11-ID of the National Synchrotron Light Source-II (NSLS-II) at Brookhaven National Laboratory. A partly coherent X-ray beam of 10 $\mu$m $\times$ 10 $\mu$m size and 9.6 keV energy impinged at a grazing-incidence angle of 0.38$^\circ$ on the sapphire substrate. An Eiger-X 4M detector (Dectris) with an individual pixel size of 75 $\mu$m was located 10.3 m away from the sample, yielding good sampling of X-ray speckles. The speckle contrast, set by experimental details, was approximately 8\%.  Using the scattering angles $\alpha_i$, $\alpha_f$, $\psi$, and sample detector distance, scattering data could be understood in terms of $q_{\parallel}$ and $q_{z}$ in reciprocal space, which roughly correspond to two directions perpendicular to the projected direction of the incident X-ray beam (i.e. the two directions are approximately parallel to the Y and Z axes in Figure \ref{fig:setup} respectively). The UHV chamber and sample holder were designed specifically to alleviate sample vibration and temperature fluctuations.  Nonetheless, at the beginning of the growth process, a gradual small (1-2 pixel in total) systematic shift of the X-ray scattering speckle pattern is observed.  This is believed to be due to small shifts in sample position/orientation associated with thermal transients associated with the start of the growth.  To remove the effect, the speckle shifts were measured using spatial correlation functions and the scattering patterns were appropriately corrected before further analysis.

InN was grown by PE-ALD on an a-plane sapphire substrate using a custom process chamber equipped with mica X-ray windows, load-lock chamber, inductively coupled plasma source, and air-cooled dry vacuum pump. The plasma source is identical to the one used by the Veeco Fiji G1 ALD system. The sapphire substrate was cleaned using ultrasonic baths of acetone, isopropanol, and deionized water and then loaded into the reactor load-lock and pumped to vacuum.  Upon transfer into the process chamber, the substrate was heated to 250$^\circ$C and prepared {\it in situ} for InN growth using sequences of H$_2$/Ar and N$_2$/Ar plasma pulses to remove contaminants and nitridate the surface \cite{rosenberg2019insitu}. The final InN film orientation is expected to be predominantly (0001) \cite{nepal2013epitaxial}, but the presence of other orientations is also likely during the early stages of growth.  A typical {\it post-facto} AFM topograph of a film grown under these conditions can be seen in Fig. 6c of Nepal {\it et al.} \cite{nepal2019understanding}. Ultrahigh purity (UHP) Ar, hydrogen (H$_2$), and nitrogen (N$_2$), further purified at point-of-use, were used as carrier and plasma gases, and UHP trimethylindium (TMI) was used as the indium precursor.  Constant Ar flows of 60 and 244 sccm were maintained through the precursor delivery line and plasma source, respectively, yielding a background pressure of 160 mTorr.  Reactive and energetic nitrogen species \cite{woodward2022influence} were generated using a 300 W N$_2$/Ar plasma for which 31 sccm N$_2$ was flowed through the plasma source.  The chamber pressure during plasma exposure was 180 mTorr.  The chamber walls were heated to prevent condensation of the precursor gases. Sample temperature was monitored by a pyrometer. 

\subsection*{Averaging and calculation of TTCF}

The speckle-averaged intensity evolution shown in Fig. \ref{fig:intensityplot} was obtained by integrating along $q_z$ = $ 0.1 - 0.2 ~\textrm{nm}^{-1}$ and smoothening.

The region of $q_{\parallel}$ = 0.2 nm$^{-1}$ $\pm$ 0.004 nm$^{-1}$ and $q_{z}$ = 0.16 nm$^{-1}$ $\pm$ 0.02 nm$^{-1}$, corresponding to a Region Of Interest (ROI) of 21 pixels along $q_{\parallel}$ and 101 pixels along $q_{z}$ on the Yoneda wing is utilized to calculate TTCF as shown in Fig. \ref{fig:TTCF}. It is calculated using Eq. \ref{eqn:twotime}, where individual intensities at two different times in an ROI are multiplied and averaged $<>$ over all pixels in the ROI, and then divided by normalization terms. Normalization terms $\left\langle I(q_{\parallel},t_1)\right\rangle$ and $\left\langle I(q_{\parallel},t_2)\right\rangle$ are obtained by applying a Savitzky–Golay filter on the ROI at $t_1$ and $t_2$ until the contrast of X-ray speckles is diminished.

\begin{figure}[h!]
\centering\includegraphics[width=9cm]{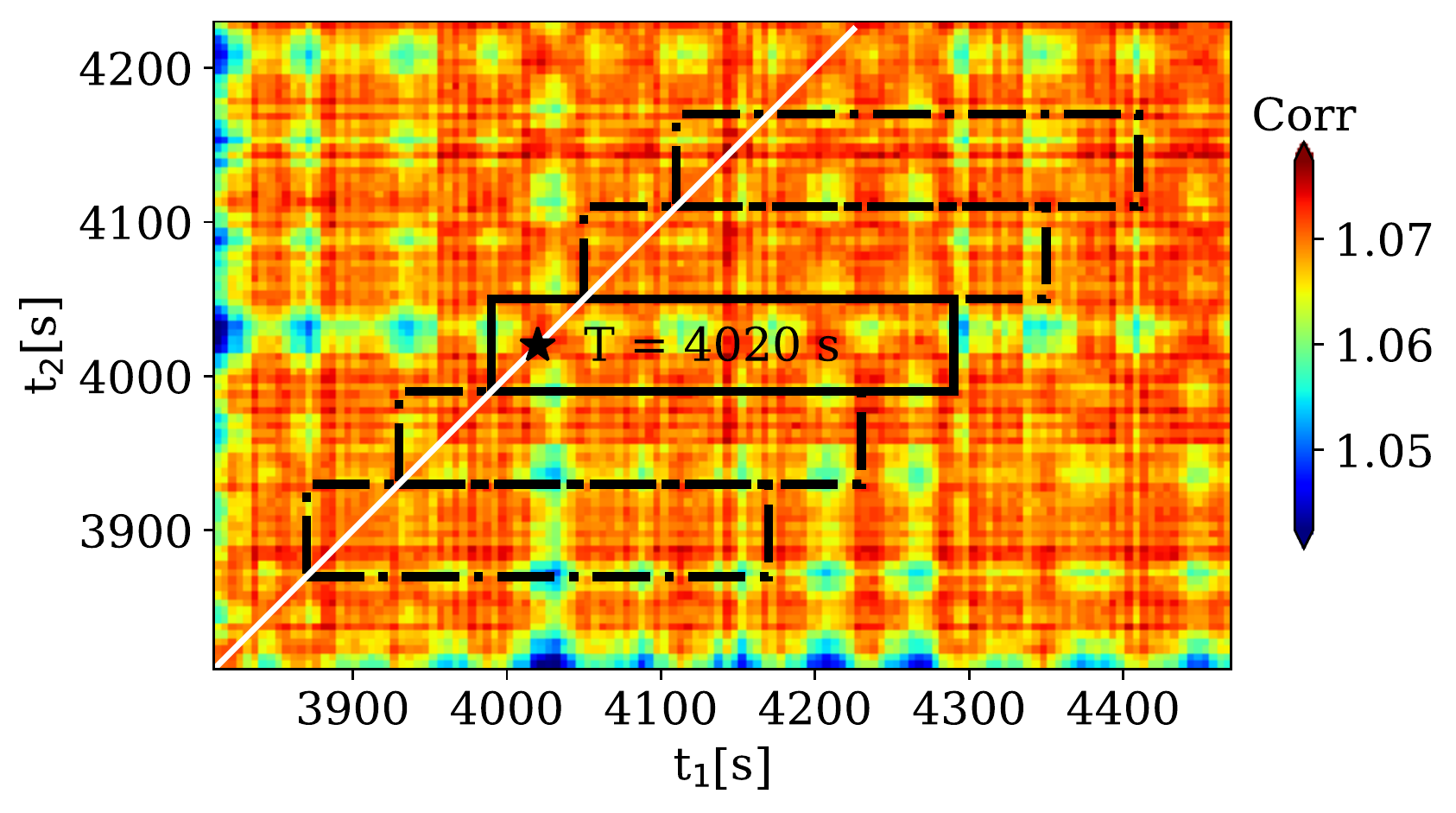}
\caption{Schematic diagram illustrating the averaging method for calculating correlation evolutions in Fig. \ref{fig:TTCF_1cycle_4cycles}}
\label{fig:averaging_TTCF}
\end{figure}

Correlation evolutions in Fig. \ref{fig:TTCF_1cycle_4cycles} were computed by averaging the TTCF over 15 growth cycles in $t_2$ for Fig. \ref{fig:TTCF_1cycle_4cycles}(a) and over 21 cycles for Fig. \ref{fig:TTCF_1cycle_4cycles}(b) and (d), with each cycle beyond the first offset in $t_1$ to always begin on the $t_1=t_2$ diagonal. Thus, $\delta t_1 = \delta t_2$ is always calculated on the diagonal.
Computed correlation evolutions were computed for an early time of $T$ = 420 s (i.e, averaged from $T$ = 0 s - 840 s or approximately the time it takes to grow $\approx$ 1.4 monolayers),  an early part of the late stage time of $T$ = 2220 s (i.e, averaged from $T$ = 1800 s - 2640 s), and a late stage time of $T$ = 4020 s (i.e, averaged from $T$ = 3600 s - 4440 s), as shown in Fig. \ref{fig:TTCF_1cycle_4cycles}. A schematic diagram portraying a late stage time of $T$ = 4020 s in TTCF averaged over 5 growth cycles in $t_2$ (instead of 15 or 21 for visual simplification purpose) is shown in Fig. \ref{fig:averaging_TTCF}.

\subsection*{Modeling}

Since no cyclic variation of the speckle-averaged intensity is observed during the late time growth, we take $\langle I_{on}({\bf q}) \rangle = \langle I_{off}({\bf q}) \rangle = \langle I_{pre}({\bf q}) \rangle = \langle I_{rel}({\bf q}) \rangle \equiv I_{ave}$, where the average brackets denote average over speckles within a given region on the detector, and $I_{ave}$ is the average intensity over that region of the detector. In addition, for simplicity we take the speckle contrasts for the four intensities to be equal; we designate them as $\left (1+\beta \right)$. The evolving speckle intensity during a growth cycle is then written as:
\begin{eqnarray}
\label{eqn:specycle}
&&\text{After Precursor Deposition:}\nonumber\\ 
&&~I_1({\bf q},\delta t)=I_{pre}({\bf q})e^{-(\delta t-\delta t_{dep})/\tau_{pre}}+I_{rel} ({\bf q}) [\, 1-e^{-(\delta t-\delta t_{dep})/\tau_{pre})} ]\,\nonumber\\
&&\text{After Plasma On:}\nonumber\\
&&~I_2({\bf q},\delta t)=I_1({\bf q},\delta t_{on})e^{-(\delta t-\delta t_{on})/\tau_{on}}+I_{on}[\, 1-e^{-(\delta t-\delta t_{off})/\tau_{on}} ]\nonumber\\
&&\text{After Plasma Off:}\nonumber\\
&&~I_3({\bf q},\delta t)=I_{on}({\bf q})e^{-(\delta t-\delta t_{off})/\tau_{off}}+I_{off}({\bf q}) [\, 1-e^{-(\delta t-\delta t_{off})/\tau_{off}} ]\,
\end{eqnarray}

Note that in this model $\langle I({\bf q},t) \rangle$ is constant in time throughout the cycle, in accord with the experiment. Also note that with the zero of $\delta t$ defined in the section above, the Plasma-Off part of the cycle wraps from the end of one cycle to the beginning of the next. %

\backmatter

\bmhead{Acknowledgments}

This work was supported by the U. S. Department of Energy (DOE) Office of Science under Grant No. DE-SC0017802. K.F.L. and P.M. were supported by the National Science Foundation (NSF) under grant no. DMR-1709380. The research conducted at the U.S. Naval Research Laboratory was supported by the Office of Naval Research (ONR). This research used the 11-ID beamline of the National Synchrotron Light Source II, a U.S. DOE Office of Science User Facility operated for the DOE Office of Science by the Brookhaven National Laboratory under Contract No. DE-SC0012704.

\noindent

\bibliography{XPCS_PEALD}%

\end{document}